\documentclass[12pt,oneside]{article}

\begin{document}
\thispagestyle{empty}
\setcounter{page}{0}
\renewcommand{\theequation}{\thesection.\arabic{equation}}

{\hfill{\tt hep-th/0201039}}

\vspace{2cm}

\begin{center}
{\bf From Big Crunch To Big Bang -- Is It Possible?}

\vspace{1.4cm}

Nathan Seiberg

\vspace{.2cm}

{\em School of Natural Sciences} \\
{\em Institute for Advanced Study} \\
{\em Princeton, NJ 08540, USA} \\
\end{center}

\vspace{-.1cm}

\centerline{{\tt seiberg@ias.edu}}

\vspace{1cm}

\centerline{ABSTRACT}

\vspace{- 4 mm}

\begin{quote}\small
We discuss the possibility of a transition from a contracting
flat space -- big crunch -- to an expanding flat space -- big
bang.  Talk given at the Francqui Colloquium, 2001, ``Strings and
Gravity: Tying the Forces Together.''

\end{quote}

\baselineskip18pt

\newpage

\setcounter{equation}{0}
\section{Introduction}

The purpose of this talk is to review the work reported in
\cite{KOSST}.  Unlike other talks which report on well understood
results, this talk should be viewed more as a {\it research
proposal}.  We will try to motivate a line of research and to
suggest a certain conjecture.

The recent exciting advances in string theory were limited to
strings in time independent backgrounds.  We know very little
about backgrounds which depend on time.  More specifically, our
understanding is limited to backgrounds which admit in asymptotic
spatial infinity a global timelike Killing vector.  This
condition rules out interesting backgrounds like those which are
important in cosmology.

Understanding cosmological solutions and especially cosmological
singularities in the context of string theory is interesting both
from a conceptual and from a pragmatic point of view.  We can
hope that through cosmology the much desired connection between
string theory and experiment can materialize.

When cosmological singularities are considered two situations
immediately come to mind:

\begin{itemize}
  \item
{\it Big bang singularity.}  Here the Universe starts from a point
and expands from it.  A commonly held point of view is that it is
meaningless to ask ``What happened before the big bang?'' because
the big bang is the beginning of time.  According to this point
of view we should understand the initial conditions of the
Universe or its wave function at the time of the big bang.
Another point of view, which appears for example in the work on
pre-big bang scenarios \cite{PBB}, holds that the Universe had an
interesting history before the big bang.  Here we will follow
this point of view, but our pre-big bang history differs
significantly from that in \cite{PBB}.

\item {\it Big crunch singularity.}  Here the Universe collapses to
a point.  This is simply the time reversal process of the big
bang singularity, and therefore if the big bang is the beginning
of time, the big crunch is the end of time.  In this case special
final conditions for the Universe or its final wave function at
the time of the big crunch have to be specified.  Having a
specific final condition independent of the details of the system
might lead to violations of causality -- no matter what the state
of the Universe is at a given time, it always ends at the same
state. Therefore, it is perhaps more intuitive to expect that
there is future beyond the big crunch.  If this is the case,
then, by time reversal, there must also be past before the big
bang.

\end{itemize}

\setcounter{equation}{0}
\section{Singularity in our Past}

In this section we will review the well known argument showing
that if the Universe expands today, it always expanded and must
have started at a singularity (our discussion is similar to that
in \cite{norev}). In other words, it is impossible for a flat
Universe like ours to undergo a period of contraction and then a
period of expansion.

Using Weyl rescaling, the Einstein term in the four dimensional
Lagrangian can be made canonical and all the terms in the
Lagrangian with at most two derivatives are
\begin{eqnarray}
{\cal L}={\cal R}-{1\over 2} G_{IJ}(\phi) \partial \phi^I\partial
\phi^J - V(\phi) \label{lagr}
\end{eqnarray}
(for simplicity we neglect gauge fields).  For homogeneous and
isotropic configurations the energy density and the pressure are
\begin{eqnarray}
  \rho=T_{00}= {1\over 2} G_{IJ}(\phi) \dot\phi^I\dot \phi^J +
  V(\phi)
\end{eqnarray}
\begin{eqnarray}
   p=-{1\over 3} g^{ij}T_{ij}= {1\over 2} G_{IJ}(\phi)
\dot\phi^I\dot \phi^J - V(\phi)   \label{rhop}
\end{eqnarray}
From these we derive that in unitary theories, where the scalar
field kinetic term is non-negative
\begin{eqnarray}
p+\rho = G_{IJ}(\phi) \dot\phi^I\dot \phi^J \ge 0
\end{eqnarray}
This is known as the null energy condition.

For metrics of the form
\begin{eqnarray}
ds^2=-dt^2+a(t)^2 \sum_i (dx^i)^2
\end{eqnarray}
with zero spatial curvature and $H \equiv {\dot a \over a} $ the
gravity equations of motion lead to
\begin{eqnarray}
\dot H = -{1\over 4} (p+\rho)\le 0
\end{eqnarray}
Therefore, reversal from contraction ($H<0$) to expansion ($H>0$)
is impossible.  (Here we assume that the time coordinate $t$
covers the whole space.) We note that this statement is similar
to the proof of the $c$ theorem in the context of the AdS/CFT
correspondence.  There it is also shown that $c$ is monotonic.

\setcounter{equation}{0}
\section{Look for a Loophole}

In this section we will look for a possible loophole in the
previous discussion.  We will show that if the system goes
through a singularity perhaps it is possible to reverse from
contraction to expansion; i.e.\ from a big crunch to a big bang.

We simplify the Lagrangian (\ref{lagr}) to a single scalar field
$\phi$.  Then we can define $\phi$ such that the kinetic term is
canonical.  For simplicity we ignore the potential and then the
Lagrangian is
\begin{eqnarray}
{\cal L}={\cal R}-{1\over 2} \partial \phi\partial \phi
\label{simlag}
\end{eqnarray}
We will consider the theory based on this Lagrangian in $d$
spacetime dimensions.

In the minisuperspace description there are three degrees of
freedom $a(\eta)$, $N(\eta)$ and $\phi(\eta)$ defined through
\begin{eqnarray}
ds^2=a^2(\eta)[-N^2(\eta) d\eta^2 + \sum_i (dx^i)^2 ]
\qquad\qquad\qquad \phi=\phi(\eta)
\end{eqnarray}
It is convenient to define the combinations
\begin{eqnarray}
a_\pm=a^{d-2\over 2} e^{\mp \gamma \phi}={1\over 2} (a_0\pm a_1)
\qquad \qquad\qquad \gamma=\sqrt{d-2 \over 8(d-1)}
\end{eqnarray}
which are constrained to satisfy  $a_0 > |a_1|$.  In terms of
these variables the Lagrangian (\ref{simlag}) becomes
\begin{eqnarray}
{\cal L} \sim { 1\over N(\eta)} \left[ -\left( { da_0 \over
d\eta}\right)^2 + \left( { da_1 \over d\eta}\right)^2 \right]
\end{eqnarray}
This theory is invariant under reparametrization  of the time
$\eta$ with an appropriate transformation of $N(\eta)$.  We can
choose a gauge $N(\eta)=1$ and impose the equation of motion of
$N$; i.e.\  $ \left( { da_0 \over d\eta}\right)^2 = \left( { da_1
\over d\eta}\right)^2$ as a constraint.  This constraint
eliminates the negative mode associated with $a_0$.  (This
negative mode is common in gravity and is familiar in the
worldsheet description of string theory.)  The coordinates $a_0$
and $a_1$ are free except that they are bounded: $a_0 > |a_1|$.

The solutions of the equations of motion and the constraint up to
shifts of $\eta$ are \cite{PBB}
\begin{eqnarray}
a(\eta) = a(1) |\eta|^{1 \over d-2} \qquad\qquad\qquad
\phi(\eta)=\phi(1) \pm {1\over 2\gamma } \log |\eta|
\label{classol}
\end{eqnarray}
$a(1)$ and $\phi(1)$ are two integration constants. We will focus
on the solutions with the plus sign.  One solution exists only
for negative $\eta$ and the other only for positive $\eta$.  The
solution with negative $\eta$ represents contraction to a
singularity with $a=0$ and $\phi=-\infty$ which happens at
$\eta=0$.  The other solution which exists for positive $\eta$
represents expansion from a singularity with $a=0$ and
$\phi=-\infty$ which takes place at $\eta=0$.  In terms of the
coordinates $a_0$ and $a_1$ the motion is free; i.e.\ linear in
$\eta$.  These variables are finite at $\eta=0$, where they reach
their bound $a_0=a_1$.

It is natural to conjecture that the two solutions can be
connected at the singularity at $\eta=0$.  Then the free
classical motion of $a_0$ and $a_1$ simply bounces off the bound
$a_0=a_1$, such that both for $\eta$ negative and for $\eta$
positive $a_0 > a_1$.  We thus conjecture that our system
contracts for negative $\eta$ and then expands for positive
$\eta$.

It is important to stress that this assumption does not violate
the no-go theorem about the impossibility of such reversal
because the system goes through a singularity.  At that point the
simple description of the dynamics in terms of the Lagrangian
(\ref{simlag}) breaks down and the no-go theorem does not apply.

In order to determine whether such reversal is possible we must
go beyond the Lagrangian (\ref{simlag}) and understand better its
microscopic origin.

\setcounter{equation}{0}
\section{Higher Dimensional Perspective}

In this section we will present a higher dimensional field
theoretic extension of the theory (\ref{simlag}) which leads to a
geometric picture of the reversal from contraction to expansion.

It is instructive to change variables to
\begin{eqnarray}
\bar g_{\mu\nu}=e^{-{4\over d-2}\gamma \phi} g_{\mu\nu}
\qquad\qquad\qquad \psi=e^{\gamma\phi}
\end{eqnarray}
In terms of these variables the action of the Lagrangian
(\ref{simlag}) is
\begin{eqnarray}
S=\int d^dx \sqrt{-\bar g} \psi^2{\cal R}(\bar g) \label{dpoac}
\end{eqnarray}
Up to rescaling of the coordinates the classical solution
(\ref{classol}) is
\begin{eqnarray}
\bar g_{\mu\nu}=\eta_{\mu\nu} \qquad\qquad\qquad \psi=\psi(1)
\sqrt{|\eta|}
\end{eqnarray}
We see that the metric $\bar g_{\mu\nu}$ is independent of
$\eta$.  In particular, it is smooth at the transition point
$\eta=0$, it is nonsingular and does not exhibit reversal from
contraction to expansion.  The solution of $\psi$ is singular at
$\eta=0$ and  since $\psi(0)=0$, the Planck scale goes to zero
there.  In these variables the singularity does not appear as a
short distance singularity (because $\bar g_{\mu\nu}$ is finite)
but as a strong coupling singularity.

We again see the need for better knowledge of the microscopic
theory at the singularity.

This theory can originate from the compactification of a $d+1$
dimensional theory on a circle or an interval.  Then $\phi$ (or
$\psi$) represents the radion -- the size of the compact
direction.  More explicitly, let the metric of the $d+1$
dimensional theory be
\begin{eqnarray}
ds^2=g^{(d+1)}_{\alpha\beta}dx^\alpha dx^\beta =\psi^4 dw^2 + \bar
g _{\mu\nu} dx^\mu dx^\nu \qquad\qquad\qquad w\sim w+1
\label{dpom}
\end{eqnarray}
We neglect the gauge field arising from the off diagonal
components of the compactification and the higher Kaluza-Klein
modes ($w$ dependence in $\psi$ and $\bar g_{\mu\nu}$).  Then,
the action (\ref{dpoac}) is simply the standard Hilbert-Einstein
action in $d+1$ dimensions of the metric
$g^{(d+1)}_{\alpha\beta}$ of (\ref{dpom})
\begin{eqnarray}
S=\int d^dxdw \sqrt{- g^{(d+1)}} {\cal R}( g^{(d+1)})
\label{dpoaca}
\end{eqnarray}
In terms of the metric $g^{(d+1)}$ up to rescaling the
coordinates our classical solution is
\begin{eqnarray}
ds^2=A^2t^2 dw^2 + \eta _{\mu\nu} dx^\mu dx^\nu
\qquad\qquad\qquad t=x^0
\end{eqnarray}
with a constant $A$.  We see here a flat $d-1$ dimensional space
times a two dimensional space ${\cal M}^2$.

The metric of ${\cal M}^2$ is
\begin{eqnarray}
ds^2=-dt^2 + A^2t^2 dw^2 = dx^+dx^- \qquad\qquad\qquad x^\pm=\pm
t e^{\pm Aw} \label{milnemet}
\end{eqnarray}
The expression in terms of $x^\pm$ shows that the space is
locally flat.  But the identification $w\sim w+1$ identified
$x^\pm \sim e^{\pm A} x^\pm $; i.e.\ the identification is by a
boost.  Without the identification this is Milne Universe which
gives a description of a quadrant of two dimensional Minkowski
space (a quadrant of the whole $x^\pm$ space).  With the
identification by a boost the space ${\cal M}^2$ is more subtle.
It has already been discussed by Horowitz and Steif \cite{HS}.

One can take various points of view about the singularity of this
space:
\begin{itemize}
  \item
Only the past cone with an end of time or only the future cone
with a beginning of time should be kept.  In terms of the
classical solutions (\ref{classol}) this corresponds to taking
one of the solutions without assuming our conjecture.
 \item
We can view the space as the quotient of two dimensional Minkowski
space by the boost. This space is not Hausdorff.  Furthermore, it
has closed timelike loops.  It is not clear whether string theory
on such an orbifold is consistent.
 \item
Our conjecture amounts to keeping only the past cone, the future
cone and the singularity that connects them.  In other words, our
space is characterized by the metric (\ref{milnemet}) with
$-\infty <t < \infty$; i.e.\ time flows from minus infinity to
plus infinity, and the singularity is a bridge from a big crunch
to a big bang.
\end{itemize}

\setcounter{equation}{0}
\section{Embedding in M Theory/String Theory}

Before embedding our geometry in string theory, we would like to
make some general comments about singularities in string theory.

For most of the singularities which have been studied in string
theory the analysis starts by taking the Planck scale to
infinity. By doing that, gravity decouples from the dynamics at
the singularity and the remaining interactions are described by a
local quantum field theory (or noncommutative field theory or
little string theory). All such singularities are characterized
by being at a finite distance in moduli space.

The singularity we are interested in here is quite different.
First, we cannot take the Planck scale to infinity, and therefore
gravity cannot be ignored. This singularity takes place at an
infinite distance in moduli space ($\phi \to -\infty$), but the
motion to the singularity involves a change in the scale factor
$a$, and hence it takes only a finite time.

It is common in string theory that singularities in its General
Relativity approximation become less singular when the stringy
corrections are taken into account. Also, it is often the case
that distinct classical spaces are connected in string theory at a
point which appears singular in the General Relativity
approximation.  The flop transition \cite{flop} and the conifold
transition \cite{conifold} are well known examples of these
phenomena. It is therefore possible that the past cone and the
future cone are also connected in a smooth way in string theory.

The conjectured two cone geometry ${\cal M}^2$ can be embedded in
string theory in many ways.  Let us mention a few of them:

\begin{itemize}
\item
M theory on $R^9\times {\cal M}^2$.  This background is flat
except at $t=0$.  Therefore it is an interesting candidate to be a
solution of the equations of motion.  Since spacetime
supersymmetry is broken, it is not obvious that $R^9\times {\cal
M}^2$ is an exact solution of the full equations of motion of M
theory.  However, for large $|t|$ the background is approximately
flat $R^{10,1}$ compactified on a large circle. Since the circle
is large, the equations of motion of eleven dimensional
supergravity give a good approximation of the full equations of
motion, and they are obviously satisfied.  Therefore, at least
for large $|t|$ this background is an approximate solution of the
equations.

We can view this background as a time dependent background for
type IIA string theory on $R^9\times R$ where $R$ is the time
coordinate in ${\cal M}^2$. This description is weakly coupled
and has small curvature for a certain range of $t$ near $t\approx
0$ (but not too close to zero) and for an appropriate range of the
parameter $A$.  There it is straightforward to see that the
leading order equations of motion of type IIA string theory are
satisfied.

\item
M theory on $R^9\times {\cal M}^2/Z_2$.  This is a $Z_2$ orbifold
of the previous example.  It shares all the features we mentioned
there.  The two branes at the end of the interval move toward
each other at negative $t$, collide at $t=0$ and bounce off each
other at positive $t$.  The entire process looks like two banging
cymbals.

\item
M theory on $R^8\times S^1\times {\cal M}^2$.  This is an $S^1$
compactification of the first example and can be interpreted as
type IIA string theory on $R^8\times {\cal M}^2$.  Here the type
IIA background is flat except at $t=0$, where the curvature is
infinite. This suggests that it could be a solution (though not a
supersymmetric one) of the classical equations of motion of string
theory. Since the string coupling can be made arbitrarily small,
the string loop corrections are small, and perhaps can be under
control.

\end{itemize}

\setcounter{equation}{0}
\section{Conclusions}

In conclusion, we conjecture a transition through a spacelike
singularity from a big crunch to a big bang. Only a detailed
stringy analysis can prove or disprove the conjecture.

If our conjecture is correct, in these models time has no
beginning or end.  In a sense this is the most conservative
assumption about the evolution of such systems; i.e.\ they are
subject to standard quantum evolution without special initial or
final conditions.

This idea (perhaps not quite this model) opens the door to the
possibility that the big bang singularity is preceded by a period
of contraction ending in a big crunch.  As in the pre-big bang
models \cite{PBB}, if the Universe existed for a long time before
the big bang, many questions in cosmology can have new solutions
and have to be revisited.

The original ekpyrotic model \cite{KOST} has a problem of
reversal from contraction to expansion \cite{linde}.  Perhaps our
banging cymbals can be incorporated into the ideas of this model
and make a viable model of early cosmology.

\vspace{5mm}

\noindent {\bf Acknowledgments}: We would like to congratulate
Marc Henneaux for well deserved recognition, and to praise the
Francqui Foundation for the good choice.  We thank them for the
invitation to the Francqui Colloquium, 2001, ``Strings and
Gravity: Tying the Forces Together,'' in Brussels and for
organizing an extremely stimulating meeting. We would also like
to thank J.~Khoury, J.~Maldacena, J.~Polchinski, B.~A.~Ovrut,
P.~J.~Steinhardt and N.~Turok for many useful discussions. This
work was supported in part by grant \#DE-FG02-90ER40542.


\vspace{5mm}


\begin{thebibliography}{99}
\bibitem{KOSST} J.~Khoury, B.~A.~Ovrut, N.~Seiberg,
P.~J.~Steinhardt and N.~Turok, hep-th/0108187.
\bibitem{PBB} For a recent review of the pre-big bang scenario,
see G.~Veneziano, hep-th/0002094.
\bibitem{norev} R.~Brustein and R.~Madden,
Phys. Lett. B {\bf 410} (1997) 110; R.~Brustein and G.~Veneziano,
Phys. Lett. B {\bf 329} (1994) 429. N.~Kaloper, R.~Madden and
K.A.~Olive, Nucl. Phys. {\bf B452} (1995) 667; Phys. Lett. B {\bf
371} (1996) 34.  R.~Easther, K.~Maeda and D.~Wands, Phys. Rev. D
{\bf 53} (1996) 4247.
\bibitem{HS} G. Horowitz and A. Steif, {\it Phys. Lett. B}{\bf 258}
(1991) 91.
\bibitem{flop} P.S.~Aspinwall, B.R.~Greene and D.R.~Morrison,
Nucl. Phys. {\bf B416} (1994) 414; E.~Witten, Nucl. Phys. {\bf
B403} (1993) 159.
\bibitem{conifold} A.~Strominger, Nucl. Phys. {\bf B451} (1995) 96;
B.R.~Greene, D.R.~Morrison and A.~Strominger, Nucl. Phys. {\bf
B451} (1995) 109.
\bibitem{KOST} J.~Khoury, B.A.~Ovrut,
P.J.~Steinhardt and N.~Turok, hep-th/0103239.
\bibitem{linde} R.~Kallosh, L.~Kofman, A.~Linde and A.~Tseytlin,
hep-th/0106241, and references therein; D.~Lyth, hep-ph/0106153;
R. Brandenberger and F. Finelli, JHEP {\bf 0111} (2001) 056
hep-th/0109004.

\end{thebibliography}
\end{document}